\documentclass[12pt,preprint]{aastex}

\newcommand{\beq}{\begin{equation}}
\newcommand{\eeq}{\end{equation}}

\def\3he{$^3$He\,}
\def\he4{$^4$He\,}

\shorttitle{Energy Partition between Electrons and Magnetic Field}
\shortauthors{C. Y. Yang \& S. M. Liu}

\begin{document}

\title{Energy Partition between Energetic Electrons and Turbulent Magnetic Field in Supernova Remnant RX J1713.7-3946}

\author{
     Chuyuan Yang\altaffilmark{1, 2, 3}
    \and
     Siming Liu\altaffilmark{1}
     }
   \altaffiltext{1}{Key Laboratory of Dark Matter and Space Astronomy,
     Purple Mountain Observatory,Chinese Academy of Sciences, Nanjing 210008, China, liusm@pmo.ac.cn}
\altaffiltext{2}{National Astronomical Observatories/Yunnan Observatory, the Chinese Academy of Sciences, Kunming 650011,   China, chyy@ynao.ac.cn}
\altaffiltext{3}{Key Laboratory for the Structure and Evolution of Celestial Objects, Chinese Academy of
     Sciences, Kunming 650011}

\begin{abstract}

Current observations of supernova remnant (SNR) RX J1713.7-3946 favor the leptonic scenario for the TeV emission, where the radio to X-ray emission is produced via the synchrotron process and the $\gamma$-ray emission is produced via the inverse Comptonization of soft background photons, and the electron distribution can be inferred from the observed $\gamma$-ray spectrum with a spectral inversion method. It is shown that the observed correlation between the X-ray and $\gamma$-ray brightness of SNR RX J1713.7-3946 can be readily explained with the assumption that the energy density of energetic electrons is proportional to that of the magnetic field in such a scenario. A 2D magnetohydrodynamic simulation is then carried out to model the overall emission spectrum. It is found that the total energy of electrons above $\sim 1$ GeV is equal to that of the magnetic field. This is the first piece of observational evidence for energy equipartition between energetic electrons and magnetic field in the downstream of strong collision-less astrophysical shocks of SNRs.

\end{abstract}

\keywords{acceleration of particles --- ISM: supernova remnants --- magnetohydrodynamics (MHD) --- radiation mechanisms: non-thermal --- shock waves --- turbulence }

\section{INTRODUCTION}
\label{intro}

It is general accepted that extended non-thermal radio astrophysical sources, such as the lobes of radio galaxies, pulsar wind nebulae, super-nova remnants (SNRs), and even some astrophysical jets, are produced through synchrotron emission of energetic electrons in a turbulent magnetized plasma. It has been argued that the energy density of transient agencies, such as turbulent magnetic field, energetic electrons and ions are in some kind of universal energy partition in a strongly turbulent collision-less plasma \citep{h98, f08, fl10, z10}. For given synchrotron radio emission, the energy equipartition between the magnetic field and energetic electrons minimizes their total energy content \citep{h04}. Direct evidence for such an energy equipartition in radio galaxies has been emerging quickly during the past decade with the availability of relatively more complete spectral data \citep{c05, hc05, ks05, ab10}.
For lobes of FR II radio galaxies, \citet{c05} first found that the magnetic field is very close to the equipartition value.
\citet{mg07} showed that the energy density of energetic electrons is higher than that of the magnetic field for a power-law distribution of the electrons and there is evidence that the magnetic field energy density decays faster than that of the electrons from hot spots at the end of the radio lobes, where both magnetic field and energetic electrons are presumably generated by collision-less astrophysical shocks. An earlier study of Pictor A by \citet{hc05} revealed similar results. Although these observations only cover the radio and X-ray bands, their results are consistent with an energy equipartition between electrons above $\sim$ 1 GeV and the magnetic field near the hot spots.

During the past decade, extensive studies have been carried out on the TeV-bright shell-type SNR RX J1713.7-3946 to explore the origin of galactic cosmic rays \citep{ah06}. Non-thermal emission dominates its emission spectrum from radio to TeV $\gamma$-rays and no thermal emission has been detected \citep{cc04}, the $\gamma$-ray luminosity is only a factor of a few lower than the synchrotron X-ray luminosity, and it has a very hard spectrum in the $\sim$ GeV energy range \citep{ab11}. These observations favor the leptonic origin for the GeV$-$TeV emission with a low mean magnetic field of $\sim 10\ \mu$Gauss \footnote{Rapid variations of X-ray filaments with a width of about 0.1 light year on a timescale of $\sim 1$ year has been interpreted as due to radiative energy loss of TeV electrons in strong ($\sim 1$ mG) fields \citep{u07}. The strong field may only exist in these filaments with a very small volume filling factor, and alternative explanations for this variability with a weak mean field have been proposed as well \citep{bu08, lf08}.
} and a well-constrained distribution of energetic electrons \citep{p06, kw08, ll11, yl11, e12}. This provides a unique opportunity to determine the energy partition between energetic electrons and the magnetic field in the downstream of collision-less astrophysical shocks of SNRs.

Although the remnant has a well-defined shell structure, the brightness profiles show significant fluctuations in both the radial and azimuthal directions \citep{ah06}, and there is a good correlation between the X-ray and $\gamma$-ray brightness, which is difficult to explain with simple models of particle acceleration \citep{ac09}.
Magnetic fields play an important role in collision-less astrophysical shocks and in the acceleration of charged particles. They can be amplified by cosmic ray induced streaming instability in the presence of efficient particle acceleration \citep[e.g.][]{b04,rs09}. In a turbulent medium, they can also be amplified by turbulent motion in the shock downstream \citep{gj07}. The latter has been explored recently by \citet{gs12} who performed extensive 2D numerical simulations for SNR blast wave interacting with a turbulent plasma background and found that the magnetic field can also be amplified by Rayleigh-Taylor convective flows induced at the contact discontinuity of the shock flows.

In this paper we first show that the observed correlation between X-ray and $\gamma$-ray brightness suggests that the energy density of the accelerated electrons is proportional to that of the magnetic field in the leptonic scenario for the $\gamma$-ray emission (Section 2). A 2D magnetohydrodynamic (MHD) simulation is then used to study the effect of magnetic inhomogeneity on the emission spectrum (Section 3). Via fitting of the overall emission spectrum of this remnant, it is found that the total energy of electrons above 1 GeV is very close to that of the magnetic field (Section 4).
Discussions and our conclusions are given in Section 5.

\section{Correlation between X-ray and $\gamma$-ray Brightness}

\begin{figure}[htb]
\centering
\includegraphics[width=8 cm]{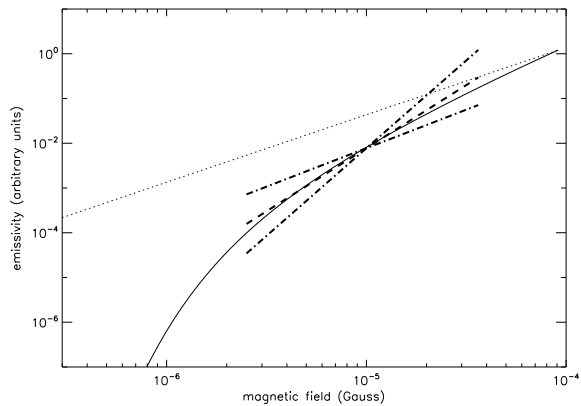}
\caption{Dependence of the synchrotron emissivity on the magnetic field for the electron distribution given by equation (\ref{ed}). The vertical axis has arbitrary units.
The solid line and dotted lines represent photon energies at 1 keV (X-ray) and 1 GHz (radio), respectively, and the two have been re-scaled for the purpose of illustration. The dotted line has a slop of 1.5 as expected. The curvature of the solid line is due to the high energy cutoff of the electron distribution.
The dashed and dot-dot-dashed lines show the correlation implied by X-ray and TeV observations.
\label{corr}}
\end{figure}

Using a spectral inversion method, \citet{ll11} showed that the overall distribution of accelerated electrons in SNR RX J1713.7-3946 can be expressed as
\begin{equation}
F_{\gamma_{e}}\sim\gamma_{e}^{-s}\exp[-(\gamma_{e}/\gamma_{max})^{\delta}],
\label{ed}
\end{equation}
where $\gamma_{e}$ is the electron Lorentz factor, $\gamma_{max}=1.3\times10^7$ is the high Lorentz factor cutoff, the spectral index $s=2$ and the cutoff index $\delta=0.6$ \citep{yl11}.
Detailed analysis of multi-wavelength observations of SNR RX J1713.7-3946 by \citet{ac09} revealed a good correlation between the X-ray and TeV brightness:
\begin{equation}
F_X \propto F_\gamma ^{2.41\pm0.55}
\label{br}
\end{equation}
where $F_X$ and $F_\gamma$ represent the X-ray and TeV brightness, respectively.
\citet{ac09} found that this correlation favors a leptonic origin for the TeV emission and is difficult to explain in simple acceleration scenarios. Magnetic fields play an essential role in the acceleration of charged particles, which is especially true in a turbulent weakly magnetized plasma, where both magnetic field and energetic particles can be produced by turbulent motions \citep{fl10}. It is therefore natural to assume a fixed energy partition between the magnetic field and the accelerated electrons in the turbulent downstream of the SNR shocks \footnote{For lobes of radio galaxies, there is evidence that this energy partition varies with the distance from the hot spots, which suggests that the coupling between the magnetic field and energy electrons is different from that in the downstream of strong collision-less shocks in SNRs \citep{k93, hc05, mg07}. The acceleration of particles should play a more dominant role in the shock downstream than in lobes of radio galaxies, where particle transport and energy loss are important.}. For the electron distribution given by equation (\ref{ed}),
Figure \ref{corr} shows the dependence of the synchrotron emissivity $\epsilon_0$ at 1 GHz (dotted line) and 1 keV (solid line) on the magnetic field for a given energy density of accelerated electrons (equation (3.39) of \citet{p70}).

If the magnetic field energy density is proportional to the energy density of the accelerated electrons, then the TeV emissivity due to the IC process is also proportional to the magnetic field energy density for whatever distribution of the external background photons \citep{p06} \footnote{For synchrotron self-Compton emission, the scaling will be non-linear in $B^2$.}. The magnetic field in Figure \ref{corr} will be proportional to the square root of the $\gamma$-ray emissivity via the IC process. The brightness is proportional to the emissivity averaged along the line-of-sight. The observed correlation between the X-ray and $\gamma$-ray brightness equation (\ref{br}) therefore implies:
\begin{equation}
 \epsilon_0\propto \epsilon_X/\langle B^2\rangle \propto \epsilon_\gamma^{2.41\pm0.55}/\langle B^2\rangle \propto \langle B^{2.82\pm1.10}\rangle,
\label{e0}
\end{equation}
where $\epsilon_X\propto F_X$ and $\epsilon_\gamma\propto F_\gamma\propto \langle B^2\rangle$ are the X-ray and $\gamma$-ray emissivity averaged along the line-of-sight, respectively, and ``$\langle\rangle$'' represents average along the line-of-sight \footnote{Strictly speaking $\epsilon_X = \langle\epsilon_0 B^2\rangle$ and $\epsilon_\gamma^{2.41\pm0.55}/\langle B^2\rangle \propto \langle B^{4.82\pm1.10}\rangle/\langle B^2\rangle$.}.
This correlation is indicated by the dashed and dot-dashed lines in Figure \ref{corr}. The observed correlation between X-ray and $\gamma$-ray brightness therefore can be readily explained with an assumed energy partition between magnetic field and accelerated electrons on the scale of the resolution of TeV images. The required mean field of $\sim 10\ \mu$G is also consistent with the leptonic scenario. On the other hand, for the radio emissivity averaged along the line-of-sight $\epsilon_R$, we have (see equation (3.50) of \citet{p70})
\begin{equation}
\epsilon_R/\langle B^2\rangle \propto \epsilon_0\propto B^{(s+1)/2}=B^{1.5}\,.
\end{equation}
Then the model predicts that the radio brightness
\begin{equation}
F_R\propto \epsilon_R\propto \langle B^{3.5}\rangle \propto\epsilon_\gamma^{1.75}\propto F_\gamma^{1.75}\,,
\label{corr1}
\end{equation}
which may be tested with better radio observations. The variance of the brightness profile therefore is smallest in the $\gamma$-ray band and biggest in the X-ray band.

\section{MHD Simulation}
\label{models}

SNR RX J1713.7-3946 was first discovered in X-rays with the
{\it ROSAT} all-sky survey in 1996 \citep{pa96}. It is close to the Galactic plane and its distance and ages are about 1 kpc and 1600 yrs, respectively \citep[e.g.][]{f03,cc04}, and the radius of the remnant is about 10 pc \citep{ac09}.
To have a more quantitative study of the emission spectrum, the dynamical evolution of an SNR shock propagating into a turbulent ambient medium is simulated with
the time-dependent ideal MHD equations of mass, momentum, and energy conservation:
\begin{equation}
\partial_t \rho+\nabla\cdot(\rho\bf u ) =0\,,
\end{equation}
\begin{equation}
\partial_t \rho{\bf u}+\nabla\cdot\left(\rho{\bf uu}-{{\bf BB}\over 4\pi}\right)+ \nabla P^\prime =0 \,,
\end{equation}
\begin{equation}
\partial_t E +\nabla\cdot\left[(E+P^\prime){\bf u}-{{\bf B(u\cdot B)}\over 4\pi}\right]=0\,,
\end{equation}
and the induction equation:
\begin{equation}
\partial_t {\bf B}+\nabla\times({\bf u\times B }) =0 \,,
\end{equation}
where $P^\prime=P+B^2/8\pi$ is the total pressure with thermal pressure $P$ and magnetic pressure $B^2/8\pi$, and cgs units have been adopted in this paper.
$E$ is the total energy density:
\begin{equation}
E=\frac{P}{\gamma-1}+\frac{1}{2}\rho u^2+\frac{B^2}{8\pi},
\end{equation}
$\rho$, $\bf u$, $\bf B$, and $\gamma=5/3$ are the plasma mass density, fluid velocity, magnetic field vector, and the adiabatic index, respectively.



We use the code PLUTO version 3.1.1 developed by \citet{m07} to solve the ideal MHD equations. A constrained transport scheme is used to enforce the $\nabla \cdot \mathbf{B}=0$ condition.
We model the simulation in a two-dimensional Cartesian coordinate ($x, y$) with $2000\times2000$ uniform  grids.  The size of simulation domain is chosen to be 40 pc $\times$ 40 pc to cover the extension of young SNRs . The supernova blast wave is driven by the injection of thermal energy and mass in a small circular region at the center of simulation box, in which the density is assumed to be constant corresponding to a plateau volume ($4\pi r_{\rm inj}^3/3$) with $r_{\rm inj}=0.5$ pc. The initial magnetic field and density in the background plasma include an average component and a turbulent component. For the sake of simplicity, we assume a constant average magnetic field $B_0$ along the $x$ direction and a constant average gas density $n_0$ following \citet{gs12}.

Both density and magnetic fluctuations are generated by the assumption of a Kolmogorov-like power-law spectrum of the form
\begin{equation}
P(k) \propto \frac{1}{1+(kL)^\Gamma},
\end{equation}
 where the spectral index $\Gamma$ depends on the dimensionality and equals to $8/3$ for 2D systems.
$k$ is the magnitude of the wavevector and $L=3\ \mathrm{pc}$ is the turbulence coherence length. The turbulence
is generated by summing a large number of discrete wave modes with random phases \citep{gj99}.
The random component of magnetic field is given by
 \begin{eqnarray}
\delta \textbf{B} (x, y) = \sum^{N_m}_{n=1} &\sqrt{C_B2\pi k_n \Delta k_n
P_B(k_n)} (\sin \theta_n \hat{x} - \cos \theta_n \hat{y})  \nonumber \\
& \times\exp (i \cos \theta_n k_n x + i \sin \theta_n k_n y + i \phi_n)
\end{eqnarray}
where $P_B(k_n)$ represents the power of the wave mode $n$ with wavenumber $k_n$. The turbulence distributes randomly in propagation direction $-1<\cos\theta_n<1$ and with a random phase $0<\phi_n<2\pi$. $C_B$ is a normalization constant determined by $\langle \delta B^2\rangle=B_0^2$, where $\langle\rangle$ represents averaging in space.

The density fluctuations satisfy the following probability distribution \citep{gj07,bl00}:
\begin{equation}
n(x,y)=n_0\exp(f_0+\delta f)
\end{equation}
where $f_0$ is a constant chosen to give the average density $n_0$ and the description of $\delta f$ is similar to the turbulent part of the magnetic field:
\begin{eqnarray}
\delta f(x, y) =&\sum^{N_m}_{n=1} \sqrt{C_f2\pi k_n \Delta k_n P_f(k_n)}  \nonumber \\
&\times\exp(i \cos \theta_n k_n x + i \sin \theta_n k_n y + i \phi_n)\,,
\end{eqnarray}
with the normalization constant $C_f$ determined by $\langle(\delta n)^2\rangle=0.4n_0^2$.

\begin{figure}[htb]
\centering
\includegraphics[width=7 cm]{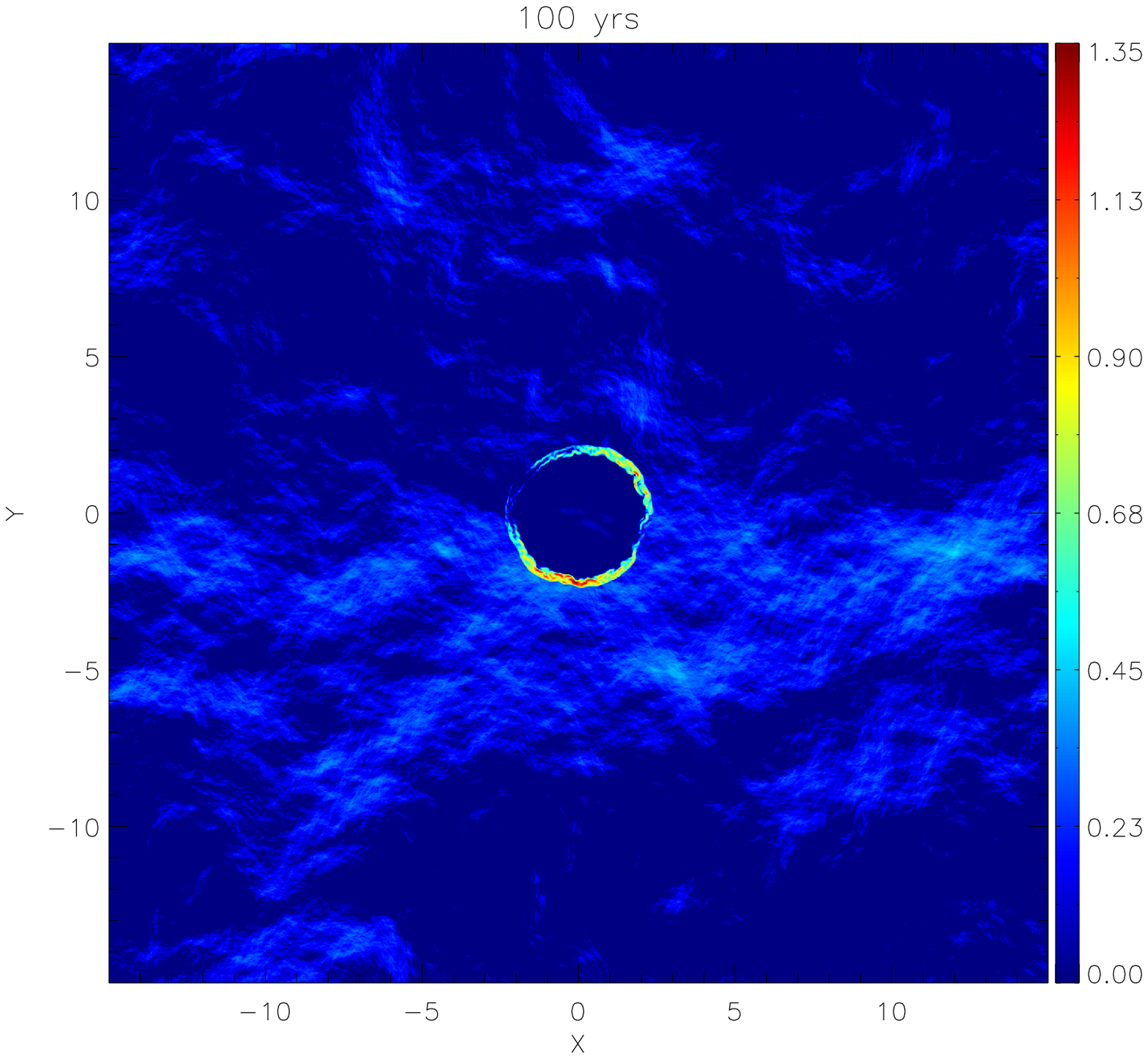}
\includegraphics[width=7 cm]{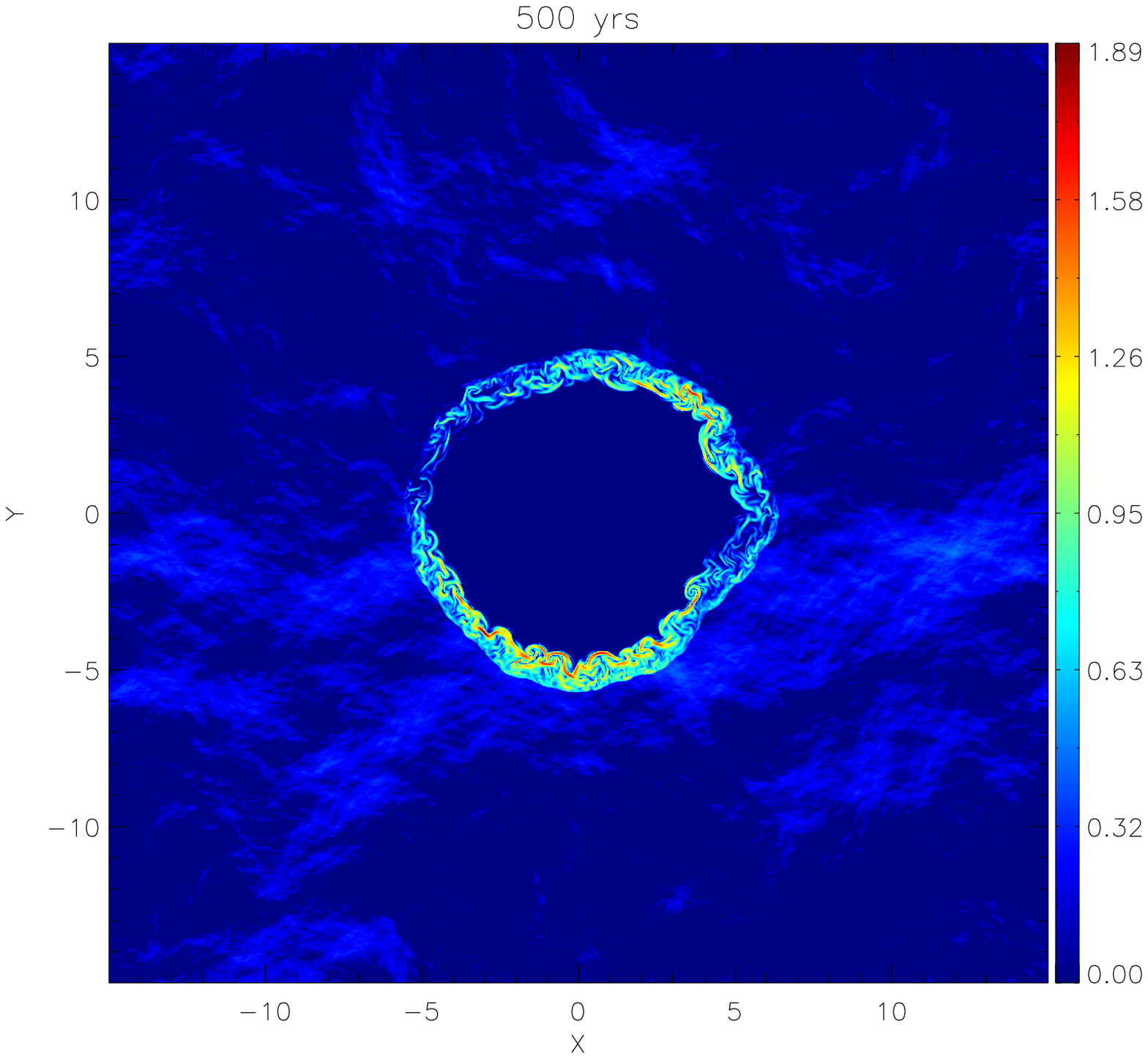}
\includegraphics[width=7 cm]{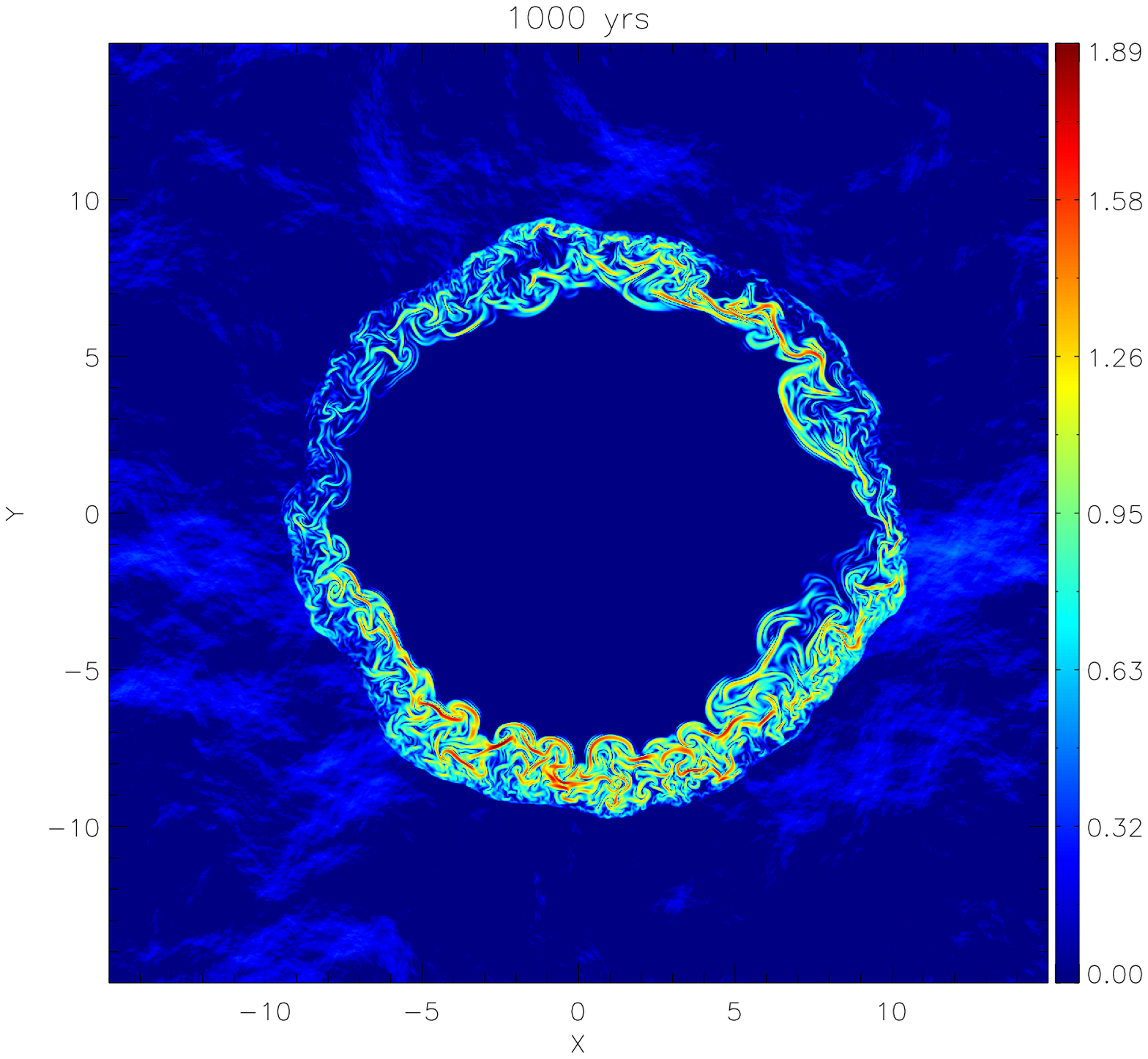}
\includegraphics[width=7 cm]{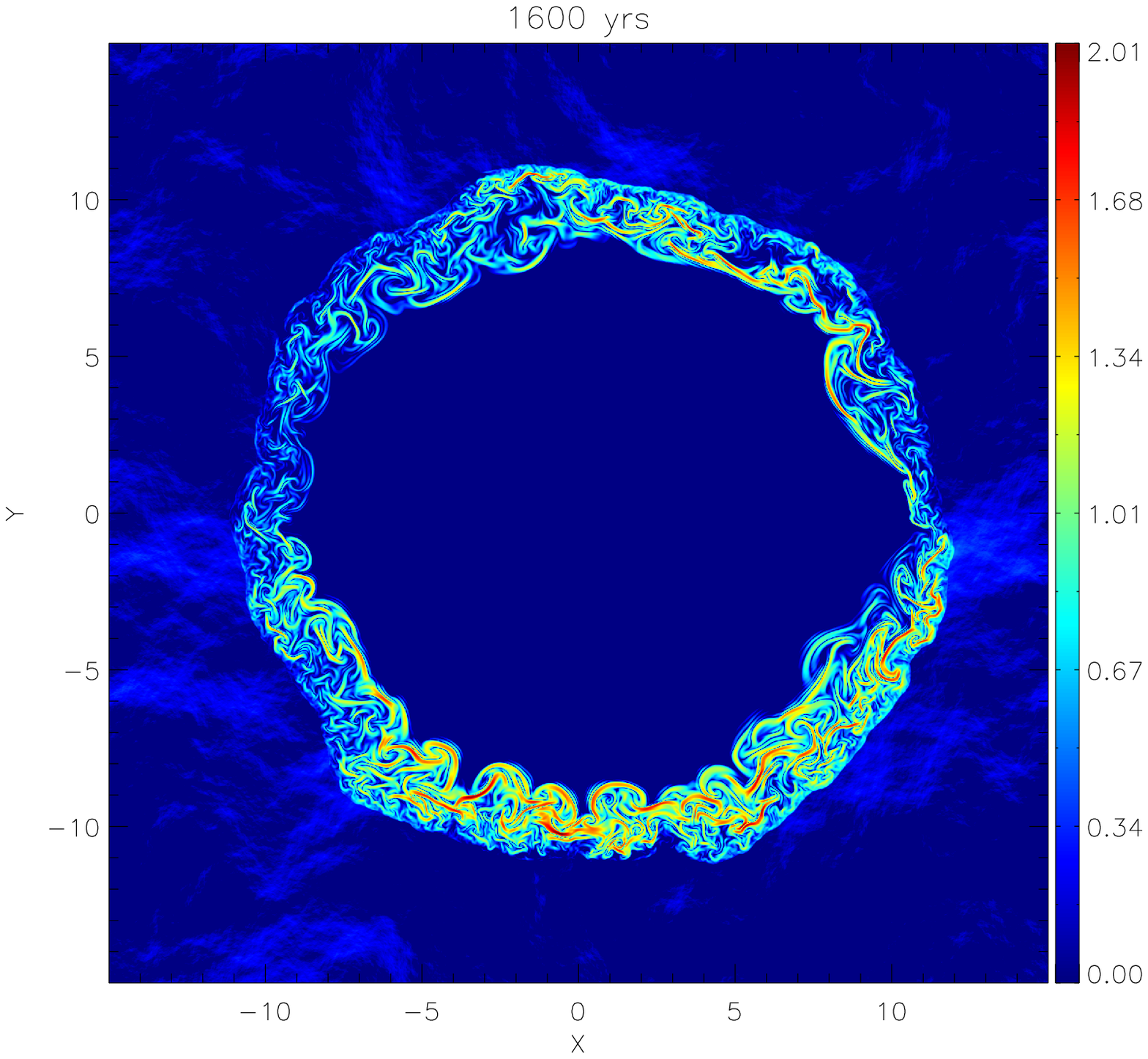}
\caption{The structure of the magnetic field  at 100, 500, 1000, and 1600 years of a run with $n_0=0.04$ cm$^{-3}$ and $B_0=0.8\ \mu$G. The color scale shows the logarithm of the magnetic field amplitude in units of $\mu$G. The rest of the model parameters are shown in the text.
\label{emag}}
\end{figure}

\citet{gs12} carried out extensive study of 2D MHD simulations of SNRs in a turbulent medium. We ran a simulation with $n_0=0.04$ cm$^{-3}$, $B_0=0.8\ \mu$G, the total injected internal energy $E_{SN}=2\times10^{51}$ ergs, and the total mass of the ejector $M_{ej}=2M_\odot$. Figure \ref{emag} shows the evolution of the magnetic field structure, which is consistent with the results of \citet{gs12}. The units of the coordinate are pc and the color scale is the logarithm of the magnetic field amplitude in units of $\mu$G. The mean magnetic field increases gradually with time and reaches a value of about $15 \mu$G by the end of the simulation at 1600 years. We actually adjusted $B_0$ to produce a mean field of $15 \mu$G at 1600 years as required to fit the overall emission spectrum \citep{ll11}. The peak magnetic field reaches about 100 $\mu$G at 1600 years. On average, the turbulent magnetic field has been amplified by a factor of $\sim 20$ by 1600 years mostly due to the Rayleigh-Taylor convective flows \citep{gs12}. Near the front of the forward shock, the field is only amplified by a factor of a few, which is consistent with earlier MHD simulations \citep{b01, gj07}.
 Due to the presence of a mean magnetic field in the $x$ direction, the magnetic field amplification in the direction perpendicular to the mean field is stronger than that along the mean field. However there are significant fluctuations no related to the large scale magnetic field as discovered by \citet{b01} with 3D simulations.
  Since the ejector is uniform, the magnetic field is very weak in the downstream of the reverse shock.

\section{Spectrum}

\begin{figure}[htb]
\centering
\includegraphics[width=6 cm]{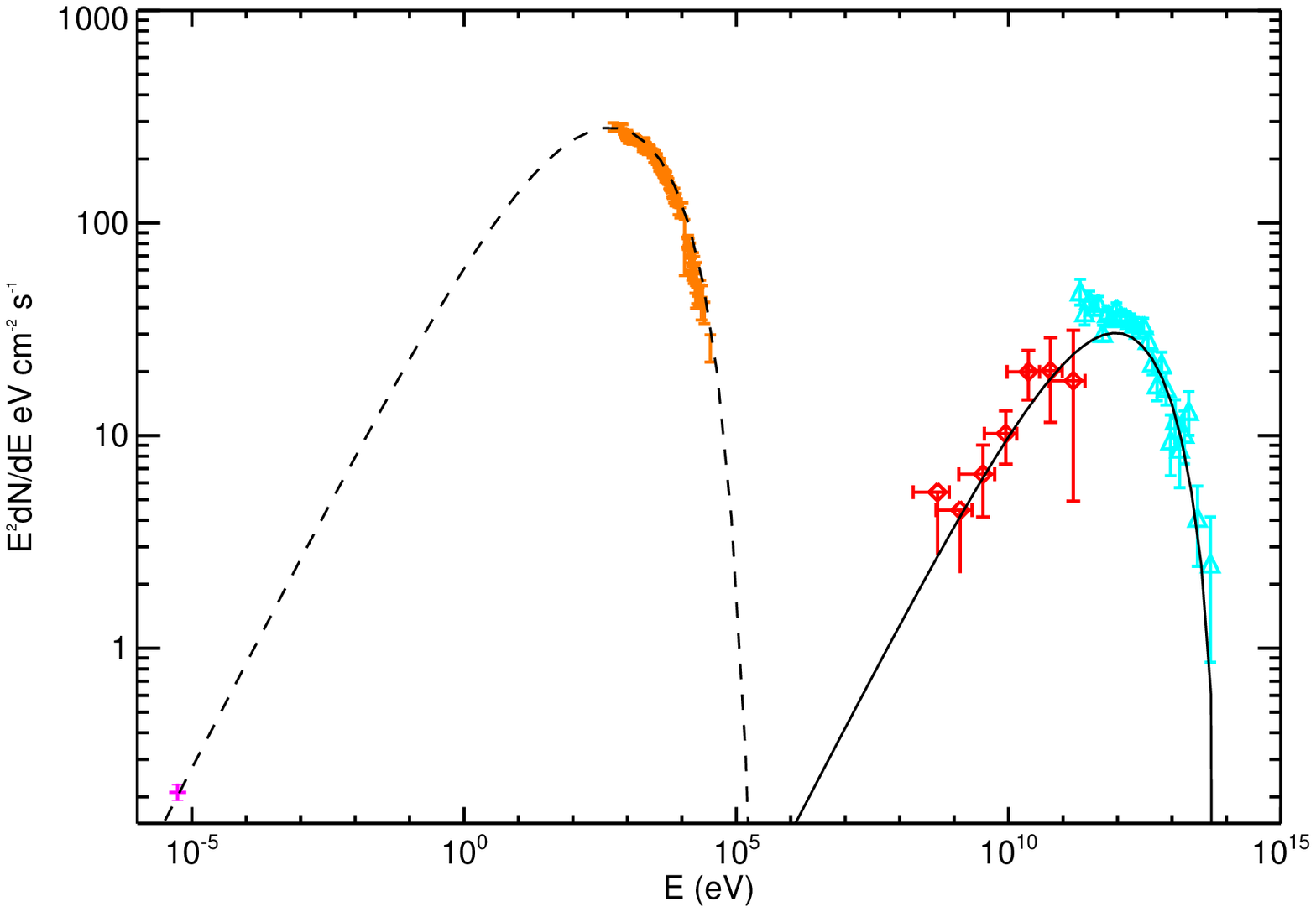}
\caption{{Spectral fit to observations of SNR RX J1713.7-3946 with a 2D MHD simulation. The different emission processes are synchrotron (dashed line), IC (solid line). The data are from \citet[][in radio]{ac09}, \citet[][Suzaku X-ray]{t08}, \citet[][Fermi-LAT]{ab11}, and \citet[][HESS]{ah06}. Note that the two lowest energy Fermi-LAT points are upper limits. The seed photon field in the IC process includes an IR component with a temperature of T = 30 K and an energy density of $1.2\ {\rm eV\ cm}^{-3}$ and the microwave background radiation \citep{ll11}. 
}
\label{spec}}
\end{figure}

To use MHD simulations to study the nonthermal radiative characteristics of SNRs, one needs to consider the acceleration of charged particles, which depends on the injection process at low energies and may also have strong kinetic effects \citep{bv10, fl10}. A self-consistent treatment of these issues in 2D case hasn't been carried out and is beyond the scope of this investigation.
On the other hand, as shown above, observations may be used to constrain the properties of accelerated particles.
In the following, 
we simply assume that the energy density of energetic electrons is proportional to the magnetic field energy density with the electron distribution in general given by:
\begin{equation}
\frac{dN(x,y, \gamma_e)}{d\gamma_e}=CB(x,y)^2 F_{\gamma_e} .
\label{ng}
\end{equation}
where $C$ is a normalization factor derived by fitting the radio to TeV spectrum of the SNR.
 Figure \ref{spec} shows a spectral fit to observations of SNR RX J1713.7-3946 with the model described above.
 The magnetic field structure at 1600 years of the 2D MHD simulation (Figure \ref{emag}) is used.

The total energy of electrons above 1 GeV and the magnetic field are $3.1\times 10^{47}$ ergs and $3.2\times 10^{47}$ ergs, respectively, which are almost equal. The energetic electrons and magnetic field therefore are in energy equipartition. This is the most important finding of the paper. Earlier numerical simulations have shown that the detailed structure of the magnetic field can be sensitive to the dimension and numerical resolution of the simulation \citep{b01, gs12}. This raises the issue whether the energy equipartition discovered here is a reliable result. We note that for a given electron distribution and background soft photon field the energy partition between the field and energetic electrons is mostly determined by the emission spectrum and the volume of the emission region both of which are well-constrained by observations \footnote{The thickness of the shell with strong magnetic fields at 1600 years in Figure \ref{emag} is compatible with the TeV image of this source \citep{ah06}. For a given IC emission, the energy density of energetic electrons is inversely proportional to the volume of the emission region.}. This is also the reason why the one-zone model can also lead to a rough energy equipartition between the field and energetic electrons \citep{f08, yl11}. The structure of the magnetic field is expected to introduce slight variation on the energy partition as far as the probability distribution of the field amplitude is not too different from a smooth broad distribution such as a log-normal one shown in the paper by \citet{gs12}. The MHD simulation here therefore just makes the model more realistic giving rise to more quantitative results. The variance of the magnetic field structure caused by the dimension and numerical resolution of the simulations is not expected to change our conclusion on the energy partition.

The inhomogeneity of the magnetic field and the assumed local energy equipartition between the field and energetic electrons can enhance the synchrotron radiation efficiency slightly. Compared with the spectral fit of one-zone emission models \citep{lf08, yl11, ll11}, the inhomogeneity of the source structure does not affect the overall spectral fit significantly. In general one may assume that the electron energy density $E_e\propto B^p$ with $0\le p\le2$. Then $p=0$ corresponds to a uniform electron distribution, which may be appropriate for some lobes of radio galaxies \citep{hc05}. In this case, the overall synchrotron emissivity should be lower than that for $p=2$ as assumed here. The relaxation of our assumed energy partition between energetic electrons and magnetic field therefore does not affect the overall emission spectrum and efficiency significantly. However, the correlation between the brightness of the synchrotron and IC components  will be different for values of $p$ different from 2. In general instead of equations (\ref{e0}) and (\ref{corr1}), we have $\epsilon_0\propto \langle B^{p(1.41\pm0.55)}\rangle$ and $F_R\propto F_\gamma^{1+1.5/p}$, respectively.


\section{Discussions and Conclusions}

X-ray and $\gamma$-ray observations of SNR RX J1713.7-3946 show that the emission region is comparable to the volume enclosed by the forward shock of the remnant \citep{ac09, ah06}. Previous studies based on a one-zone emission model have shown that the total energy of energetic electrons above $\sim 1$ GeV and the magnetic field within the remnant is comparable in the leptonic scenario \citep{f08, yl11}. Through an MHD simulation, we show here that electrons above $\sim$ 1 GeV is in local energy equipartition with the magnetic field in the downstream of the forward shock, which also naturally explains the observed correlation between the X-ray and $\gamma$-ray brightness. This result may be readily applied to other similar remnants dominated by nonthermal emission to explore their nature \citep{ylb12}. The inhomogeneity of the magnetic field in combination with the assumed energy partition between energetic electrons and the magnetic field may increase the synchrotron emission efficiency of energetic electrons slightly but does not affect the overall emission spectrum.
These conclusions also weakly depend on the properties of the assumed soft background photons and the detailed structure of the emission region and may be valid in other similar astrophysical contexts such as in the hot spots of radio galaxies \citep{mg07}.

The model predicts that the radio brightness is proportional to the 1.75th power of the $\gamma$-ray brightness, which can be tested with future observations. If the model is further validated, it will imply dominance of the electron acceleration via stochastic scattering with a magnetized turbulent plasma instead of by diffusive shock acceleration because most of the magnetic field is generated via the Rayleigh-Taylor instability near the contact discontinuity and turbulent motion in the shock downstream. This local energy equipartition addresses the injection problem of the electron acceleration. In the context of stochastic particle acceleration, we expect that the electron distribution has a low energy cutoff or spectral break at the proton rest mass energy $\sim $1 GeV  due to resonant interactions with whistler waves at even lower energies at the acceleration site \citep{pl04, l06}. To complete the phenomenological theory of stochastic electron acceleration, one also needs to understand what determines the spectral index above $\sim 1$ GeV and the high-energy cutoff of the electron distribution \citep{fl10}. In this paper, we assume that these two parameters are the same through the remnant, which implies a harder X-ray spectrum in brighter region. X-ray observations by \citet{ac09} (see their Fig. 7) appear to show the opposite. The remnant expands faster toward the southeast direction, where the X-ray spectrum is also harder. The model therefore can be further improved by considering the dependence of the electron spectral index and high energy cutoff on the shock speed.

One should note that the observed correlation between the X-ray and TeV brightness may be viewed as evidence for energy equipartition on scales greater than the resolution of the TeV images. There is no guarantee that such a local energy equipartition remains valid at even smaller scales. TeV observations with better spatial resolution are needed to check the validity of energy equipartition between energetic electrons and magnetic field on even smaller scales in SNRs. We also note that observations of lobes of radio galaxies clearly show variation of energy partition between energy electrons and magnetic field across the lobes \citep{hc05}. The coupling between energetic electrons and magnetic field in these lobes is likely different from that in the downstream of strong collisionless shocks, where both energetic particles and magnetic field are generated efficiently \citep{k93}. There is also evidence for a spectral break in the distribution of relativistic electrons below $\sim 1$ GeV \citep{ab10}. Caution must be exercised when applying the local energy equipartition between energetic electrons above $1$ GeV and magnetic field to other nonthermal radio sources.

Finally when using the 2D MHD simulation to model the observed brightness, it is assumed that the line-of-sight averaged quantities have similar statistical properties as those derived directly from the 2D simulation. A 3D simulation is needed to clarify this issue \citep{b01}.

\acknowledgements

We thank Drs. Hui Li and Fan Guo for helpful discussions and for providing a code to generate turbulent ISM, and the referee Prof. Hardcastle for a very informative report which also motivates our discussion on the relaxation of the assumed energy partition between magnetic field and energy electrons.  This work is partially supported by NSFC grants: 11163006, 11173064, 11233001, and 11233008.

{}

\end{document}